\documentclass[aps,twocolumn]{revtex4}
\usepackage{graphicx,color}

\begin{document}
\title{Quantitative laboratory observations of internal wave reflection on ascending slopes}

\author{Louis Gostiaux$^1$, Thierry Dauxois$^1$, Henri Didelle$^2$,  Joel
  Sommeria$^2$,  Samuel Viboud$^2$} \date{\today} 
\affiliation{1. Laboratoire
  de Physique, UMR-CNRS 5672, ENS Lyon, 46
  All\'{e}e d'Italie, 69364 Lyon c\'{e}dex 07, France\\
  \email{Louis.Gostiaux@ens-lyon.fr,Thierry.Dauxois@ens-lyon.fr}\\
  2. Laboratoire des \'Ecoulements Géophysiques et
Industriels (LEGI), UMR 5519 CNRS-UJF-INPG,
21 rue des Martyrs, 38000 Grenoble, France}
\bibliographystyle{plain}

\begin{abstract}
Internal waves propagate obliquely through a stratified
  fluid with an angle that is fixed with respect to gravity. Upon
  reflection on a sloping bed, striking phenomena are expected to
  occur close to the slope.  We present here laboratory observations
  at moderately large Reynolds number.  A particle image
  velocimetry (PIV) technique is used to provide time resolved
  velocity fields in large volumes.  The generation of the second and
  third harmonic frequencies are clearly demonstrated in the impact zone.
  The mechanism for nonlinear
  wavelength selection is also discussed. Evanescent waves with
  frequency larger than the Brunt-V\"ais\"al\"a frequency are detected
  and experimental results agree very well with theoretical
  predictions. The amplitude of the different harmonics after
  reflection are also obtained.
Hydrodynamic waves. 
\end{abstract}
  \maketitle

\section{Introduction}

The oblique propagation of internal waves follows from the dispersion
relation that monochromatic perturbations of frequency $\omega$ have
to satisfy
\begin{equation}
\omega=\pm N\sin\theta,
\label{dispersionrelation}
\end{equation}
where $N$ is the Brunt-V\"ais\"al\"a frequency
\begin{equation}N=\sqrt{-\frac{g}{\rho_0}\frac{\partial\rho}{\partial z}},\end{equation}
$g$ being the gravity, $\rho(z)$ the ambient density profile and
$\rho_0$  a reference density. This dispersion relation shows that
for a fixed frequency, the direction in which energy propagates
with respect to the horizontal, $\theta$, is fixed. Moreover,
Eq.~(\ref{dispersionrelation}) determines that phase and energy
propagate in perpendicular directions. 
For set-up with Brunt-V\"ais\"al\"a frequency $N$  independent of $z$,
observations of internal
waves have invariably showed~\cite{MR67,DDF04} this transverse and
oblique propagation.

The above dispersion relation is obtained, away from any turbulent portions of the domain, by substituting a plane
wave $A\exp[ik(x\sin\theta+z\cos\theta)]$ of wavenumber $k$ and
amplitude $A$ in the wave equation governing the horizontal
velocity $u$ 
\begin{equation}\label{eqhyperbolic}
u_{ttxx}+u_{ttzz}=-N^2u_{xx},
\end{equation}
where subscripts denote partial derivative, $x$ and $z$ refer to
Cartesian coordinates and $t$ to time.   The vertical velocity
$w$, the perturbation pressure $p$, the streamfunction $\psi$ and
the perturbation density satisfy the same hyperbolic
equation~(\ref{eqhyperbolic}).

The presence of a solid horizontal bottom, a free surface or an
oblique slope results in a reflected wave with the same intrinsic
wave frequency as the incident one. {\em Linear} internal wave
reflection on a sloping bottom has been treated analytically by
different authors~\cite{P77,E85,T87}. The striking consequence of
the geometric focusing of linear internal waves has also been
reported~\cite{ML95,MBSL97,M01}. Under appropriate conditions, it
leads to internal wave attractors in confined stably
stratified fluids.
For {\em critically} incident waves for which the slope and energy
propagation angles coincide, the linear inviscid analysis becomes
singular and an infinite amplitude of the reflected wave was
predicted~\cite{P77}. 

Experiments on internal wave reflection were
first performed by Cacchione and Wunsch~\cite{CW74} using conductivity
probe measurements. A tidal-like excitation was generated in a 5 m
long tank by a horizontally oscillating paddle; the incident and
reflected waves from a $15^\circ$ slope were separated using
periodogram estimates to compute wave amplitude and wavenumber.
Although these pioneering results were of poor quality
(compared to what is possible nowadays), their shadowgraph experiments
showed nevertheless striking microstructures along the slope,
reminiscent of an array of vortices along the slope's boundary layer.
Thorpe and Haines~\cite{TH87} measured dye band displacements on a
$20^\circ$ slope providing qualitative agreement with linear theory
and also noticed three dimensional boundary layer structures.
Subsequently, Ivey and No\-kes~\cite{IN89} estimated the mixing
efficiency above a $30^\circ$ slope submitted to modal excitation and
visualized with rainbow Schlieren technique: the weakening of the
background stratification was measured and the corresponding change in
potential energy compared to the mechanical work provided by the
wave-maker. The internal wave reflection mechanism has also been
studied in close connection with its importance for the creation of
nepholoid layers by McPhee and Kunze~\cite{MK02}. More recently,
Dauxois, Didier and Falcon~\cite{DDF04} performed Schlieren
experiments on critical reflection, focusing on the boundary layer
upwelling events and providing good qualitative agreement with the
previous weakly nonlinear study~\cite{DY99}.  Finally, using synthetic
Schlieren measurements, Peacock and Tabaei~\cite{PT05}
showed the second harmonic generation.  To our
knowledge, no local quantitative amplitude measurements of internal
wave reflection have been so far performed.

Recently, the {\em weakly nonlinear} theoretical issue has been put
forward. It has been shown by Dauxois and Young~\cite{DY99} that
the singularity can be healed using matched asymptotic expansion.
Their analysis describes the build up
of the reflected wave along the slope for a incident plane wave. 
Always from the theoretical point of view, the reflection of internal 
waves was revisited by Tabaei, Akylas and Lamb~\cite{TAL05} for the case of
a narrow incident beam but for non-critical angles. In this case, they have predicted
the generation of harmonics in the steady regime; this result has not been 
addressed experimentally yet. Besides, fully nonlinear numerical
simulations have also been performed to examine the behavior of
large-ampli\-tu\-de internal gravity waves impinging on a
slope~\cite{SR98,JIA99,S99,JIA00,ZS01}.

Here we present the results of laboratory experiments in which a
{beam} of internal waves is reflected on an oblique slope.
Experiments were carried out in the 13m diameter Coriolis
platform, in Grenoble, filled with sal\-ted water.  The large
scale of the facility allows us to strongly reduce
 the viscous dissipation along wave propagation and, moreover, quantitative results are
obtained thanks to high resolution Particle Image Velocimetry (PIV)
measurements.

The paper is organized as follows. In
Sec.~\ref{experiemnt}, we present the experimental set up and
discuss all control parameters. In Sec.~\ref{experimentalresults},
we show the experimental results. We explain in Sec.~\ref{spectralanalyiss} the spectral
analysis used to distinguish the different harmonics.  In the
following section~\ref{mechanism}, we discuss the mechanism of
wavelength selection. Section~\ref{evanescent} is devoted to the
evanescent waves, while Sec.~\ref{amplitudemeasurements} discuss amplitude
measurements. Section~\ref{conclusion} concludes and gives some
perspectives.

\section{Experiments}
\label{experiemnt}

\subsection{Experimental setup}
\label{experimentasetup}

\begin{figure}[!ht]
\centering
\includegraphics[width=8.5truecm]{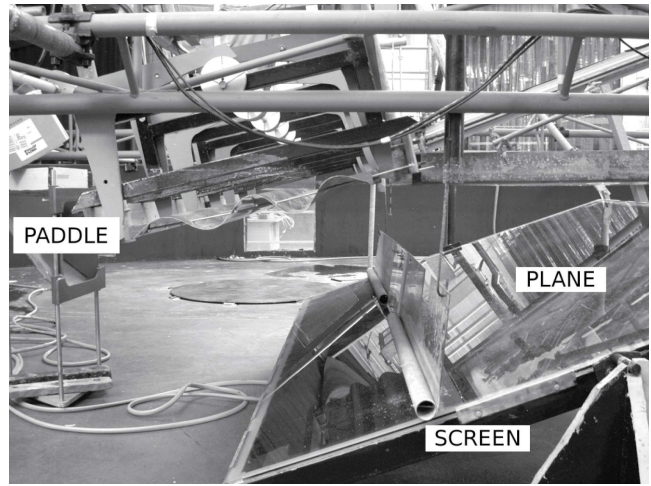}
\caption{
  Picture of the experimental setup. The 3 meters long oscillating
  paddle generates  incident internal waves which are
  impinging on the inclined glass plane visible on the left.
  One also clearly sees the screen which avoids the propagation
  toward the glass plane of harmonics also generated by the paddle.
  This picture was taken before the 13 m diameter tank has been filled
  with 1 m of stratified salt water.  }\label{fig:photo}
\end{figure}

We developed an original internal wave exciter inspired by Ivey et
al. \cite{IWS00} in order to produce a two and a half wavelengths beam. A
PVC sheet was compressed on both sides by seven squared arms
 (Fig.~\ref{fig:photo})
 fixed on a long central axis
via eccentric wheels.  The axis was set in rotation by a
stepping motor, generating a longitudinal oscillation motion
of 8 cm amplitude along the 60 cm width of the paddle. The
frequency of excitation, proportional to the angular speed of the
motor could be precisely monitored in order to vary the angle of
propagation of internal waves. The wavelength was also varied from
11.3 to 12.6 cm. The paddle itself was inclined at an angle
$\Phi=13^\circ$ with the horizontal to increase excitation
efficiency. If the main part of the energy is indeed transmitted
to a beam of internal wave of frequency $\omega$, higher harmonics
$n\omega$ are also excited by the oscillating paddle.
Nevertheless, as their frequencies are higher, their propagation
angles are also larger. Taking advantage of this fact, a screen
has been appropriately located above the bottom-end of the glass
plate so that all harmonics are reflected to the left, and do not
perturb the region of interest. 
The emitted plane wave hits a $2\times3$ meters glass plane (see Fig.~\ref{fig:photo})
, back-painted in black to
avoid parasite laser beam reflections. 

\begin{figure}[!ht]
\centering
\includegraphics[width=8.5cm]{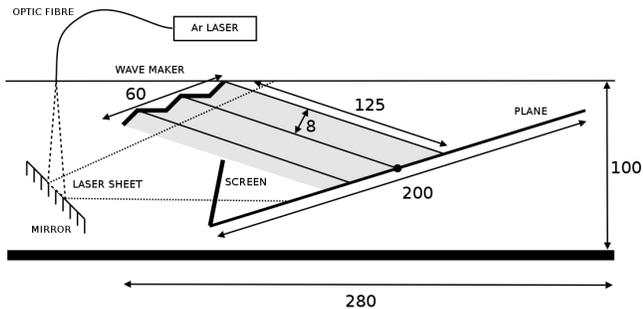}
\caption{Sketch of the vertical cross-section
of the experimental setup. The sinusoidal thick line on the top
left indicates the position of the oscillating paddle, while the
grey region defines the incident beam impinging on the sloping
glass plane. The optic fibre, indicated by the solid line on the left,
shines a laser light which  is reflected on the underwater 45
degree mirror, making a thin vertical laser sheet necessary for
the PIV measurements. Finally, the dashed line shows the position
of the screen which avoids the propagation toward the glass plate
of harmonics also generated by the paddle. A slit in the screen allows 
the laser sheet to pass through. All distances are in
centimeter.}\label{fig:side}
\end{figure}

The above experimental set up was put in the 13 m diameter
Coriolis tank filled from below with salt water and stratified by
computer controlled volumetric pumps from two 75 m$^3$ tanks, one
filled with salt water and the other one with pure water.  A fast
conductivity probe and a temperature probe were lowered slowly
into the tank using a controlled vertical micro-stepping motor to
measure the stratification. The density probe was calibrated using
an Anton Paar densitometer accurate to 0.0001 kg/m$^3$ and
0.01$^\circ$C. The linearity of the resulting density gradient
was of very good
accuracy, and only the upper 5 cm and the bottom
10 cm were not linearly stratified.
 %A typical resulting density gradient is shown in
%Fig.~\ref{fig:density}. Its linearity is seen to be of very good
%accuracy. One can notice that only the upper 5 cm and the bottom
%10 cm are not linearly stratified.
 Observations have thus been
performed in the intermediate region where the Brunt-V\"ais\"al\"a
frequency is a very well defined constant.

The stratification of 2.2\% over the 1 meter depth of the tank led to
a Brunt-V\"ais\"al\"a period $2\pi N^{-1}=13.6$s.
All experiments we discussed were performed without rotation of
the tank, and thus involve pure internal waves.

%\begin{figure}[!ht]
%\centering
%\includegraphics[width=8truecm]{Fig3GostiauxPoF.eps}
%\caption{Density versus the depth of the tank. The symbols
%  corresponds to the measurements obtained with the conductivity probe
%  while the solid line is a linear fit defining the Brunt-V\"ais\"al\"a
%  frequency. }\label{fig:density}
%\end{figure}

We used the Particle Image Velocimetry (PIV) facility of the Coriolis
Platform to obtain top and side views of the velocity field.  The
fluid was seeded with 400 microns diameter particles polystyrene
beads that were carefully prepared by a process of cooking which
decreases slightly the density and successive density separations.
One thus obtains a flat distribution of densities matching that of
the salt stratification. This process ensures that there are equal
number densities of particles at each depth. A surfactant was
added to prevent the polystyrene beads from clustering. It has
been shown that the relaxation time for particles to attain
velocity equilibrium~\cite{PFS05} is about 0.02s which is much
shorter than the characteristic time scale of the flow. As usual,
the particles are considered as passive tracers of the fluid
motion.

With this method, motion is vizualized by illuminating particles
with a laser sheet, which are followed by a digital camera. The
laser is intentionally kept out of focus (approximately 1 cm sheet
width), enabling tracking of particles despite some cross sheet
displacement. Velocity fields within the plane of the laser sheet
are obtained by comparing patterns in two subsequent image frames
(taken 1 s apart).

The 6 Watt green Argon laser was placed above the free surface,
but the light was reflected on a 45 degree mirror placed in the
water (see Fig.~\ref{fig:side}). Thanks to a second underwater 45
degree mirror, the images were acquired by a 1024*1024 pixels CCD
camera also located above the free surface.   The
cross-correlation PIV algorithm designed by Fincham and
Delerce~\cite{FD00} was used to convert the images into vector
fields (stored in the standard file format NetCDF).  This
algorithm provides good anti-aliasing and peak-locking rejection
procedures. Our resolution went to subpixels distortions,
corresponding to submillimetric displacements of the fluid.
Typical maximal displacement in an image pair corresponds to 5
pixels, with a measurement precision of 0.2 pixel for an
individual field (4$\%$ relative precision). This is mostly a
random error, so the precision on averaged fields is higher.
Finally, the analysis of the .nc files was performed with Matlab
software.

Typical experimental runs lasted about 20 minutes. The first 10 periods
were considered as an initial transient.  Data were thus only recorded
throughout the second stage during which the steady regime was
attained.

\subsection{Characteristics of the incident wave beam}

The set of chosen excitation periods (see
Table~\ref{table:conditions}) leads through the dispersion
relation~(\ref{dispersionrelation}) to different angles of
propagation $\theta$ ranging from 13 to $25^\circ$.  As the slope
angle $\alpha$ is set to 22$^\circ$, these cases allow us to
analyze subcritical ($\theta<\alpha$), supercritical
($\theta>\alpha$) or critical ($\theta\simeq\alpha$) reflections.
Finally, as the wavemaker is slightly tilted ($13^{\circ}$) from
the horizontal, one has to take into account that the wavelength
slightly varied from one run to another, around a typical value of
$\lambda_0=12$ cm.

\begin{table}[!ht]
\centering
\begin{tabular}{|l|l|l|l|l|l|}
\hline
Run&1&2&3&4&5\\
\hline
$T$ (s)&60&49&41.5&36&32\\
\hline
$\theta$ (deg.)&13&16&19&22&25\\
\hline
$\alpha$ (deg.)&22&22&22&22&22\\
\hline
$\lambda$ (cm)&11.3&11.7&12.0&12.2&12.6\\
\hline
$u_{max}$ (cm/s) & 0.412 &   0.566   & 0.775 &   1.000  &  1.225\\
\hline
\end{tabular}
\caption{Summary of experimental runs with all control parameters.
$T$ is the excitation period, $\theta$ the angle of energy propagation,
 $\alpha$ the angle of the slope defined
in Fig.~\ref{fig:side3} and $\lambda$ the wavelength of the
incident beam. $u_{max}$ is the maximum horizontal velocity that
has been measured.}\label{table:conditions}
\end{table}

Previous experimental studies of internal wave reflection have
reported the importance of viscous
dissipation~\cite{DDF04,PT05,GGFD05} to explain the observed
steady-state solution. Here, by contrast, the large scale of the
experiment allowed to work at a large Reynolds number.
By taking the wavelength and the velocity amplitude of the incident internal wave beam,
one gets for the Reynolds number $Re\simeq
100$, which suggests that viscous dissipation might be negligible
during the propagation toward the slope of the beam itself. This
is what can be verified in Figs.~\ref{fig:cont} where one has
used a synchronous detection-like idea. We have filtered the PIV signal 
at the excitation frequency $\omega$ (see Sec.~\ref{spectralanalyiss} 
for additional details). As the energy is expected to obliquely propagate 
with an angle $\theta$ with respect to the horizontal, the relevant quantities
are the  velocity fields
 $v_s(s,\sigma)=u\cos\theta-w\sin\theta$ and
$v_\sigma(s,\sigma)=u\sin\theta+w\cos\theta$. Both have been
measured and the first one is reproduced with false color in panel (a). The
longitudinal section presented in panel (b) with a solid line emphasizes that the
amplitude of the incident beam is constant. The dashed line which shows the velocity 
field $v_\sigma$ along the longitudinal section of panel (a) is as expected vanishingly small. 
This is an important
necessary condition to discuss quantitatively the reflection
process. Moreover, the picture emphasizes that incident phase
planes, which correspond to the same color, are parallel to the
direction of the incident group velocity, the latter being
indicated by the dashed line. This is a clear demonstration of the
orthogonality of the group velocity and the wave vector.  Finally,
panel (c) presents $v_s$ along the solid line of panel~(a)
and reveals that the width of the beam contains  two
wavelengths. This is an important point to draw a comparison with
theoretical predictions derived for plane waves~\cite{DY99}. Again, the dashed line attests that the 
velocity field $v_\sigma$ is vanishingly small.

\begin{figure}[!ht]
\centering
\includegraphics[width=8cm]{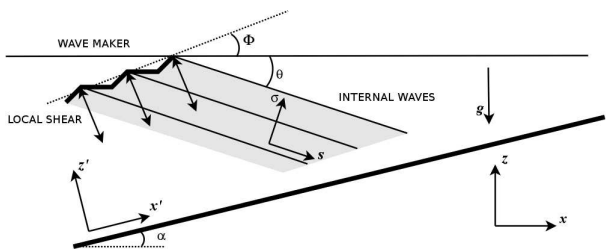}
\caption{Definition of variables.
  Three different coordinates systems will be successively used: the
  gravity-oriented one $(x,z)$, the slope-oriented one $(x',z')$ and
  finally the incident beam oriented one ($s, \sigma)$.  The PVC sheet
  compressed periodically along the dotted line provides a shear flow
  orthogonal to this line, and thus tilted at a constant angle $\Phi$
  with the vertical. This flow induces waves propagating at an angle
  $\theta$ with the horizontal, determined by the dispersion
  relation~(\ref{dispersionrelation}) while $\alpha$ is the slope
  angle.  }\label{fig:side3}
\end{figure}

\begin{figure}[!ht]
\centering
\includegraphics[width=8.5cm]{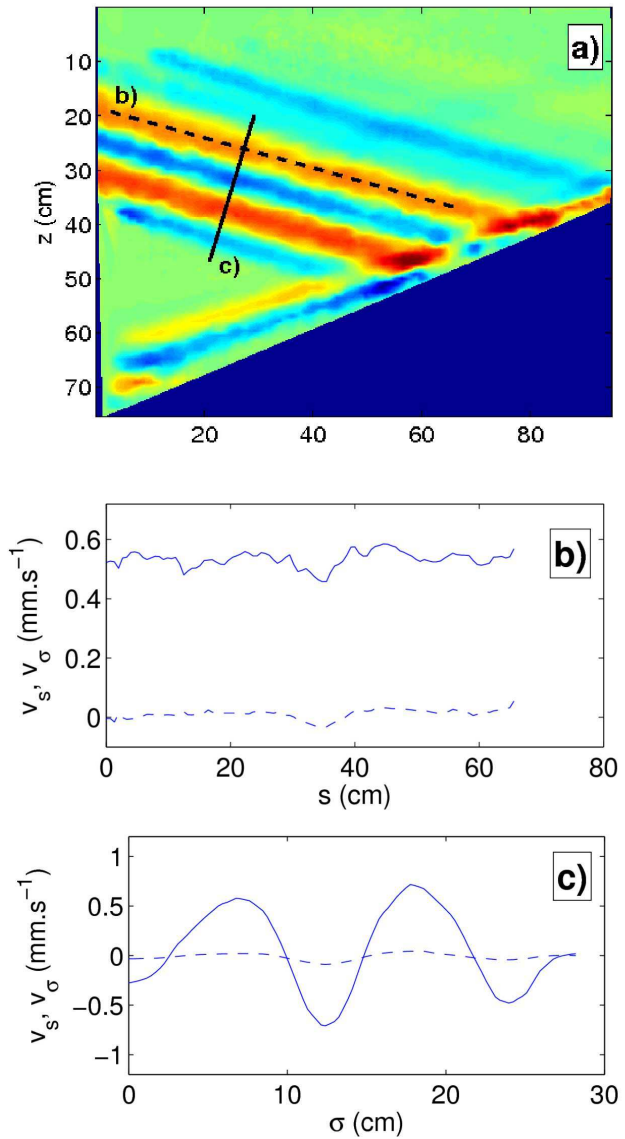}
%\includegraphics[width=8.5cm]{Fig5aGostiauxPoF}
%\vskip2mm
%\includegraphics[width=8.5cm]{Fig5bGostiauxPoF}
%\vskip2mm
%\includegraphics[width=8.5cm]{Fig5cGostiauxPoF}
\caption{{\em Picture in color.} False-color
  pattern of the along-beam ${v_s}(x,z)$ velocity for
  $\theta=16^\circ$ and $\alpha=22^\circ$ in a vertical $(x,z)$
  section. The shaded triangle corresponds to the region below the
  glass plate, unperturbed by the internal waves. The maximum velocity
  in this figure is 0.07 cm.s$^{-1}$. Panels (b) and (c) show the along-beam and cross-beam
  sections respectively, indicated by the dashed and solid lines in panel
  (a).  The longitudinal variation $v_s$ is shown in panel (b) and (c) with solid lines,
 while  the dashed lines correspond to the transversal variation $v_\sigma$. The longitudinal coordinate~$s$ and the
  transversal one $\sigma$ are defined in Fig.~\ref{fig:side3}.
}\label{fig:cont}
\end{figure}

\section{Experimental results}
\label{experimentalresults}

\subsection{Spectral analysis}
\label{spectralanalyiss}

\begin{figure}[!h]
\centering
\includegraphics[width=6.cm]{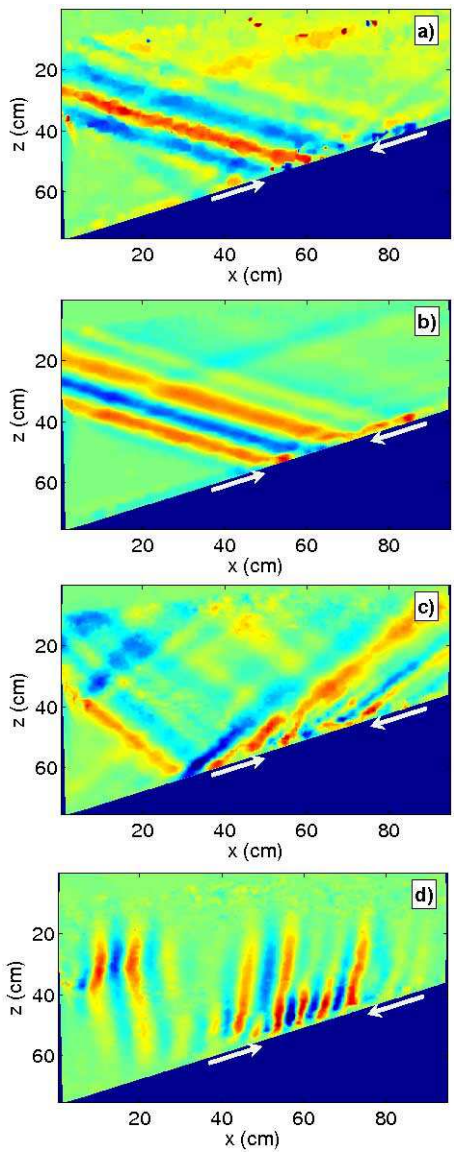}
%\vskip2mm
%\includegraphics[width=6.85cm]{Fig6bGostiauxPoF}
%\vskip2mm
%\includegraphics[width=6.85cm]{Fig6cGostiauxPoF}
%\vskip2mm
%\includegraphics[width=6.85cm]{Fig6dGostiauxPoF}
%\centering
%{\includegraphics[width=8cm]{Fig6aGostiauxPoF}
%\hskip2mm
%\includegraphics[width=8cm]{Fig6bGostiauxPoF}}
%\vskip2mm
%\includegraphics[width=8cm]{Fig6cGostiauxPoF}
%\hskip2mm
%\includegraphics[width=8cm]{Fig6dGostiauxPoF}
\caption{{\em All panels in color.} False-color velocity pattern in the vertical $(x,z)$-plane
  for the critical run 4 (see Table~\ref{table:conditions}).  Panel
  (a) presents the instantaneous horizontal velocity field $u(x,z,t)$
  while panel (b) shows the phase-averaged velocity $\langle u\rangle_1$. Panels (c) and (d) show
  respectively the second $\langle w\rangle_2$ and third $\langle
  w\rangle_3$ harmonics of the vertical velocity, $\langle
  w\rangle_n$, in the case of run 4 (see
  Table~\ref{table:conditions}).  The two white arrows define the impinging region
of the incident beam. In
panels~(c) and~(d), the rays at the left of this region should thus not be
taken into account:  they have been generated by the screen. The maximum velocity in panels (a)
  and (b) is 2 mm.s$^{-1}$ and,  in panels (c) and (d), 0.4 mm.s$^{-1}$.
}\label{fig:harmos}
\end{figure}

An important characteristic of the reflection of internal waves is the
generation of different harmonics, theoretically predicted a long time
ago~\cite{T87,DY99}, but only very recently experimentally
observed~\cite{PT05}. However the amplitudes of the different
harmonics can be very different and, consequently, hardly
distinguishable even though their propagation angles are different. A
typical example is presented in Fig.~\ref{fig:harmos}a.

This is the reason why we developed a Fourier temporal analysis of
the results, with filtering at the fundamental and higher
harmonics frequencies. Given the two components of the velocity
field provided by the PIV analysis, we compute the filtered
velocity fields. For the horizontal one $u(x,y,t)$, one thus
defines the different quantities
\begin{eqnarray}
 \langle u\rangle_n
&=&\frac{2}{t_1-t_0}\int_{t_0}^{t_1}\!\!\!\!u(x,z,t)\cos(n\omega
t+\phi)\, \mathrm{d}t.\label{fig:defu}
\end{eqnarray}
In the time interval $[t_0,t_1]$, the amplitude of the $n$-th
harmonic $\langle u\rangle_n$ is of course a function of the
spatial variables $x$ and $z$, but also of a constant phase
$\phi$, chosen with respect to the flat position of the sinusoidal
paddle. This procedure is equivalent to band pass filtering the PIV time series at each point in the domain.

Panels (b), (c) and (d) of Fig.~\ref{fig:harmos} show that this procedure is
an excellent tool to distinguish the different harmonics. Indeed, for
this almost critical reflection case (run 4, $\alpha=22^\circ$,
$\theta=22\pm1^\circ$), it is clearly apparent in panel (b), that the
reflected beam at frequency $\omega$ is absent except a small
slightly supercritical alongslope ray. Nevertheless, panels (c) and (d) show the
emitted second and third harmonics propagating with steeper
angles.  The angles of propagation with respect to the horizontal for
the second harmonic beam is $\theta_2=48\pm 1^\circ$ in agreement with
the theoretical value
\begin{equation}
\theta_n=\sin^{-1}(n\sin\theta),
\end{equation}
for $n=2$. The third harmonic is evanescent; its characteristics
will be discussed in Sec.~\ref{evanescent}. Both harmonics are
generated in the finite domain region, adjacent to the bottom, where the incident beam hits the slope.
One should not take into account the few rays located
at the left of the arrows (see
Figs.~\ref{fig:harmos}c and~\ref{fig:harmos}d): they were
generated by the oscillations of the screen. The emitted second harmonic
between the two arrows can still be spatially distinguished from these artifacts. It is important to
emphasize that the color scales differ by a factor 5 between the
two first panels and the two last ones.

In summary, Figs.~\ref{fig:harmos} exemplified that, even though
the second and third harmonics are almost invisible from the
instantaneous velocity field, they are very clearly apparent after
the filtering procedure. This filtering method is consequently
appropriate even when the amplitude is very small. To our
knowledge, even if it was previously predicted~\cite{T87,TAL05}
this is the first case where the third harmonic has been observed,
and even more important, this is the first report of experimental
quantitative measurements.

It is important to notice that the vertical velocity field is
presented in panels (c) and (d), rather than the horizontal one as
in the first two panels. Indeed, as the second and third harmonics
propagation angles are much steeper, the horizontal velocity field
is of lower quality.

An alternative possibility to keep the spatial conformation of the
reflection process while distinguishing the different harmonics is
to consider the specific kinetic energy density field of each
harmonic which can be deduced from Eq.~(\ref{fig:defu}) as
\begin{equation}
\langle E\rangle_n(x,y,\phi)=\frac{1}{2}\left[\langle u\rangle_n
^2+\langle w\rangle_n ^2\right].
\end{equation}

\begin{figure}[!ht]
\centering
\includegraphics[width=8.2cm]{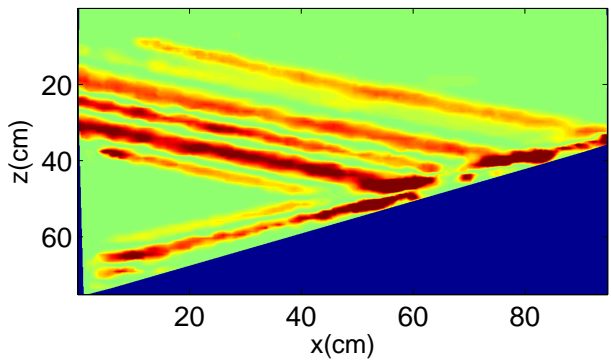}
\caption{{\em Color online.} False-color energy pattern. The
  specific kinetic energy density of the first harmonic, $\langle
  E\rangle_1$, is shown  in the vertical
  $(x,z)$-plane for the subcritical run 2 (see
  Table~\ref{table:conditions}). The maximum
  value is 5.$10^{-3}$ cm$^2$.s$^{-2}$.}\label{fig:energy}
\end{figure}

An example is shown in Fig.~\ref{fig:energy} for the subcritical
run~2. One clearly distinguishes the incident beam impinging on
the slope, and reflected downslope. It is clear that such energy
plots give less contrasted results than the velocity ones as in
Figs.~\ref{fig:harmos}. They are nevertheless extremely useful for
amplitude estimates along cross sections of the beam as discussed
below.

\subsection{Mechanism of wavelength selection}
\label{mechanism}

The spectral decomposition presented in the previous section allows
precise wavelength measurements across the different beams, even in
the impact zone. As  the reflection surface is expected to play a key
role in wavelength selection via boundary effects, we have focused our
study in the boundary region along the slope.

\begin{figure}[!ht]
\centering\includegraphics[width=8.5cm]{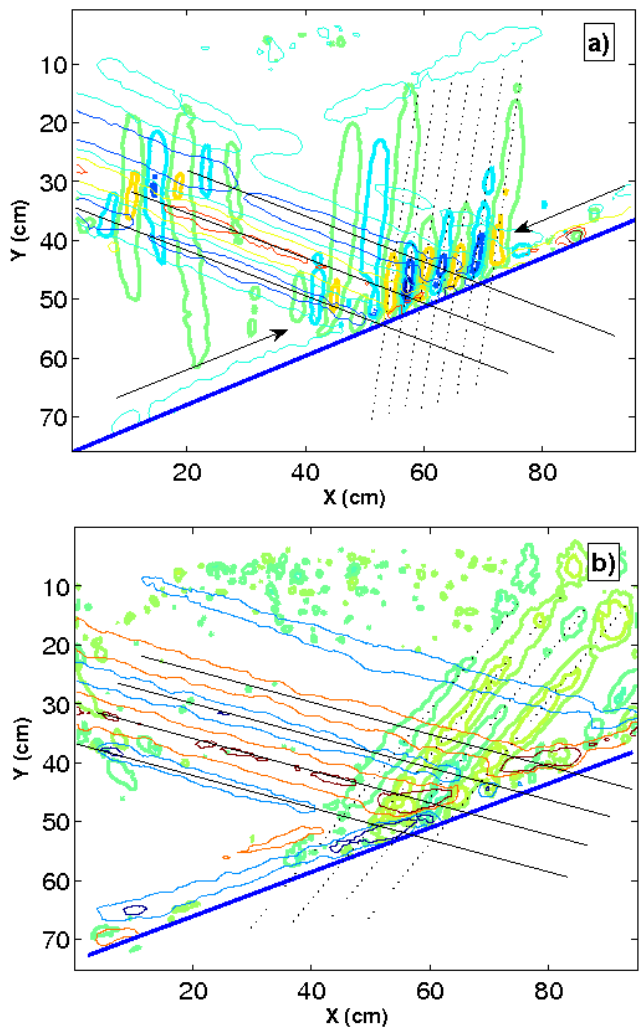}
%\centering\includegraphics[width=8.5cm]{Fig8aGostiauxPoF}
%\centering\includegraphics[width=8.5cm]{Fig8bGostiauxPoF}
\caption{{\em Pictures in color.} Superposition of the vertical velocities contourplots of the incident
  first harmonic (normal) and of the emitted third harmonic (bold).
Panel (a) presents the
  critical case (run 4) while panel (b) shows the
  {\em subcritical} case (run 2).  The solid lines show the phase lines of the
  incident first harmonic $\langle w\rangle_1$ while the dashed ones
  correspond to the phase lines of the emitted third harmonic $\langle
  w\rangle_3$. The two arrows in panel (a) indicate precisely the alongslope cross-section
  used in Fig.~\ref{fig:freq}.}\label{fig:superp}
\end{figure}

In Fig.~\ref{fig:superp}(a), the contourplots of the filtered first
harmonic (incident) and of the third harmonic (emitted) are
superimposed in order to show the alongslope wavelength selection.  It
is clearly visible that the distance between the emitted phase lines
has been strongly reduced compared to the incident ones. We emphasize
that the wavelength selection mechanism appears to occur along the
slope, where the superposition of the incident and the reflected beams
generates nonlinear interactions. Assuming that this inner region plays a
key role in the reflection of internal waves~\cite{DY99}, the only
relevant dynamical behavior of the wave field has to be taken at
$z'=0$ where the incident and the emitted waves can both be written as
$\psi(z'=0)=A \exp[i (k_{x'}x'-\omega t)]$, the amplitude $A$ being
different for the incident and the reflected waves. Nonlinear
interactions may lead to second or third harmonics terms such as
$\psi_2(z'=0)=A_2 \exp[2i (k_{x'}x'-\omega t)]$ or $\psi_3(z'=0)=A
\exp[3i (k_{x'}x'-\omega t)]$. The {\em alongslope} wavelength of the
third harmonic is thus reduced by a factor three, as highlighted in
Fig.~\ref{fig:superp}(a) by the phase lines intersections with the slope.

%\subsubsection{Supercritical cases}

\begin{figure}[!ht]
\centering
\includegraphics[width=8.75cm]{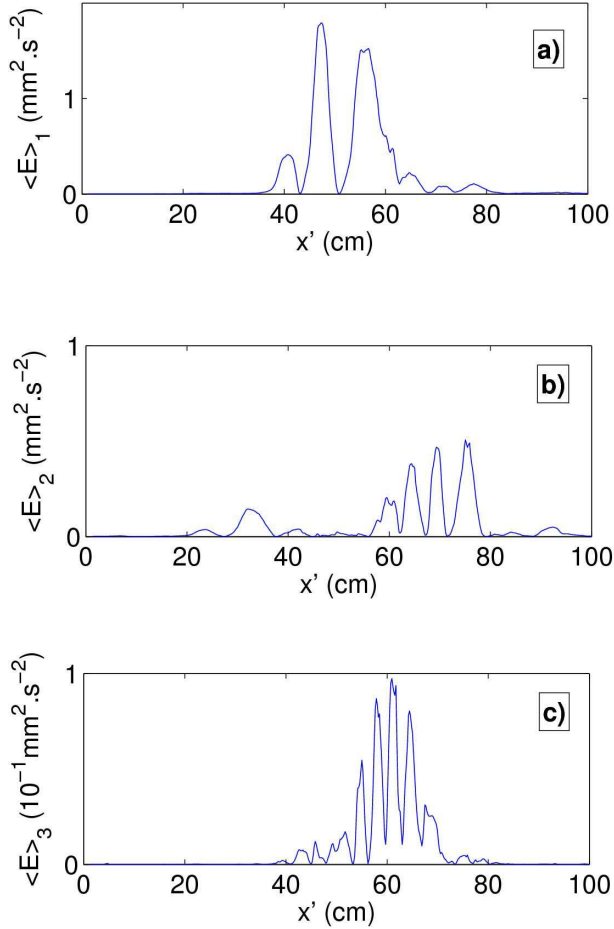}
\caption{Alongslope sections of the energy of the incident first
  harmonic (panel a), second harmonic (panel b) and of the emitted third harmonic (panel c) in
  the supercritical case (run 5).  The sections were obtained at 7 cm
  from the slope, as indicated by the arrows in
  Fig.~\ref{fig:superp}(a).  Note the difference in units for the
  ordinates.}\label{fig:freq}
\end{figure}

To gain further insight, Figs.~\ref{fig:freq} pre\-sent
quantitative measurements of this effect in the {\em
supercritical} case 5. The cross sections of the first three
harmonics at a fixed distance from the slope are presented. The
difference in $x'$-location are of pure geometrical origin and
decrease, of course, when the cross-section is taken closer to the
slope.  These pictures are analogous to the theoretical results
presented by Tabaei, Akylas and Lamb in Figs.~9(a-d) of
Ref.~\cite{TAL05}.

Let us stress that the above quantitative comparisons are possible
despite the large difference between the incident and emitted
amplitudes.  The automatic elimination of wrong vectors in the
measurements by the high quality cross-correlation PIV algorithm
(designed by Fincham and Delerce~\cite{FD00}) is the key point
here to achieve this goal. The measurements are indeed
sufficiently precise to provide meaningful results even
when the relative energy ratio between successive harmonics is
approximately 0.1. This last ratio is of course directly linked to
the parameter $\varepsilon$, used  in Refs.~\cite{DY99,TAL05} for
the small amplitude asymptotic expansion to describe theoretically
the reflection process. This is consequently not a limiting
factor.

\begin{figure}[!ht]
\centering
\includegraphics[width=8.5cm]{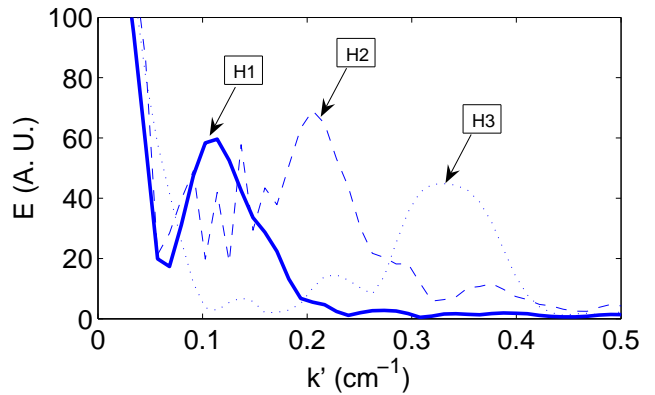}
\caption{Spatial Fourier transform of the alongslope sections
  presented in Figs.~\ref{fig:freq}. The solid line corresponds to
  $<E>_1$ (Fig.~\ref{fig:freq}(a)), the dashed one to $<E>_2$
(Fig.~\ref{fig:freq}(b)) and the
  dotted one to $<E>_3$ (Fig.~\ref{fig:freq}(c)).
This picture emphasizes the
  frequency tripling.  }\label{fig:Fourierfreq}
\end{figure}

The nonlinear wavelength selection can be clearly high\-ligh\-ted
using above the three pictures. Indeed, Fig.~\ref{fig:Fourierfreq}
presents their spatial Fou\-rier transform and clearly emphasizes
that the wavelength of the $n$-th harmonic is $n\lambda$,
explaining the wavevector tripling visible in
Fig.~\ref{fig:superp}(a) as theoretically predicted by Tabaei, Akylas
and Lamb~\cite{TAL05}.

%\subsubsection{Subcritical cases}

Experimental results in subcritical cases have unexpectedly revealed
an apparent different mechanism for the wavelength selection, when the
slope angle $\alpha$ is larger than $\theta$, the angle of energy
propagation. A typical example is shown in Fig.~\ref{fig:superp}(b).
The second harmonic has disappeared and the wavelength of the third
harmonic is equal to the incident wave when it is projected along the
slope. This possibility was already mentioned by Thorpe~\cite{T87}
since third order nonlinear interaction along the slope ($z'=0$) may
lead to third harmonics of the form $\psi_3(z'=0)=A \exp[i
(k_{x'}x'-3\omega t)]$, where the temporal frequency is tripled while
the spatial frequency is kept constant.
This is to our knowledge the first
experimental evidence of this possible nonlinear interaction. The
transition from supercritical (third harmonics alongslope wavevector
tripling and second harmonics alongslope wavevector doubling) to
subcritical case (third harmonics alongslope wavevector conservation
and absence of second harmonics), is visible in Fig.~\ref{fig:superp}(a) where for the 
critical case, both wavelengths are present at the frequency $3\omega$ (same alongslope 
wavelength on the left of the impact zone, tripled on the right). However, this 
transition remains unexplained.

\subsection{Evanescent harmonics}
\label{evanescent}

\begin{figure}[!ht]
\centering
\includegraphics[width=8.5cm]{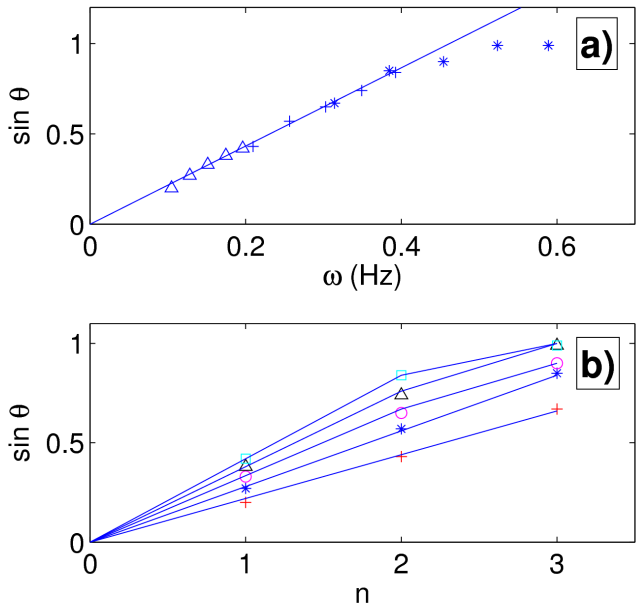}
\caption{Angular dispersion for the different harmonics. Panel (a)
presents the sine of the different angles of propagations versus
the wave frequency $\omega$. The different symbols corresponds to
the first (triangles), second (plus) and third harmonic (stars). The
solid line corresponds to the
theoretical law $\omega=N\sin\theta$ for $2\pi N^{-1}=13.6s$. Panel
(b) presents the same data as a function of the number of the
harmonics for runs 1 to 5, respectively plus, stars, circles, triangles
and squares). Error bars are smaller than the symbol size.}\label{fig:dispersion}
\end{figure}

 The angles of propagation of the different
harmonics have been measured on the filtered patterns. All the
results are listed in Table~\ref{tab:angles} for the five runs of
Table~\ref{table:conditions}. The values are also plotted in
Fig.~\ref{fig:dispersion} as a function of the pulsation $\omega$
and of the harmonic number $n$. Panel (a) attests that the first
and second harmonic are in perfect agreement with what is
theoretically expected: all corresponding symbols are on the
straight line $\omega=N\sin\theta$. This is not the case for the
third harmonic in the three last runs. When the sine of the
propagation angle is plotted versus the number of the harmonic as
in panel (b), one also sees that the third harmonic is not aligned
with the first two ones  for runs 3 to 5.

\begin{table}[!ht]
\centering
\begin{tabular}{|l|l|l|l|l|l|}
\hline
Run&1&2&3&4&5\\
\hline
$\theta_1$  (deg.) &12&16.5&19.5&22.5&25\\
\hline
$\theta_2$  (deg.) &26&35&41&48&54\\
\hline
$\theta_3$  (deg.) &43&52&?65&85&81.5\\
\hline
$\sin\theta_1$ &0.21&0.28&0.33&0.38&0.42\\
\hline
$\sin\theta_2$ &0.44&0.57&0.66&0.74&0.81\\
\hline
$\sin\theta_3$ &0.68&0.79&0.91&1.00&0.99\\
\hline
\end{tabular}
\caption{Experimentally measured angles of propagation for the
three different harmonics. The precision of the measures is
smaller than half a degree. The characteristics of the five runs
are given in Table~\ref{table:conditions}.}\label{tab:angles}
\end{table}

\begin{figure}[!ht]
\centering
\includegraphics[width=8.5cm]{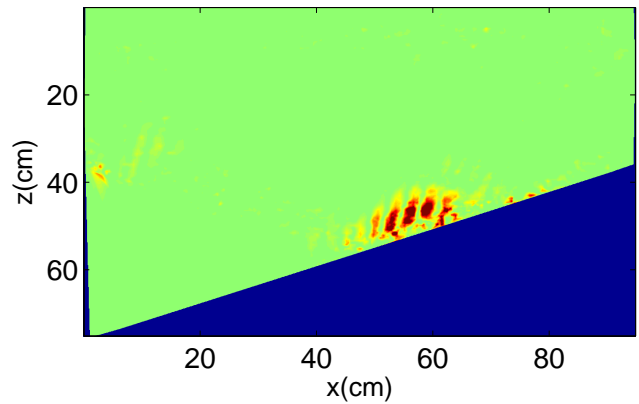}
\caption{{\em Color online.} False-color velocity pattern. The vertical third
harmonic $<w>_3$ is shown in the $(x,z)$-plane for the run 5. The
maximum velocity is 0.4 mm.s$^{-1}$.} \label{fig:trapped}
\end{figure}

As discussed for example by Tabaei et al~\cite{TAL05}, the maximum
incident angle $\theta$ for which the $n$-th harmonic can
propagate is $\sin^{-1}(1/n)$.  For larger angles, the
corresponding harmonic will be evanescent since the frequency
would be larger than the Brunt-V\"ais\"al\"a top band frequency.
For $\theta>\sin^{-1}(1/3)\simeq 19^\circ$, the third harmonic is
thus found to be evanescent : this is the case for runs 3, 4 and 5
of Table~\ref{tab:angles}, explaining the three symbols not on the
line in Fig.~\ref{fig:dispersion}(a). In these cases, the third
harmonic generated in the impact zone cannot propagate, and is
thus trapped along the slope. This is what is shown in
Fig.~\ref{fig:trapped}. Evanescent modes were also found experimentally~\cite{TII97}
in a different context, nonlinear non-resonant interaction between two internal wave rays.

If the evanescence of the wave is clarified, an angle of propagation,
{\em not vertical}, can be measured as shown for run~5 in
Fig.~\ref{fig:trapped}. Another case is also visible in
Fig.~\ref{fig:harmos}(d). It is possible to theoretically explain this
angle as follows.

\begin{figure}[!ht]
\centering
\includegraphics[width=8.5cm]{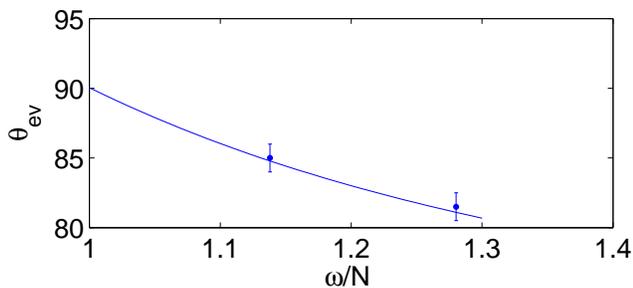}
\caption{Angle of propagation of the evanescent third harmonic for
run 4 and run 5.}\label{fig:theorie}
\end{figure}

Considering the linear equation~(\ref{eqhyperbolic}) valid for
internal waves within the Boussinesq approximation,
let us look for streamfunction  solutions $\psi$, evanescent in
the $z'$-direction, i.~e. orthogonally to the slope, but
propagating in the $x$ and $z$ direction (see Fig.~\ref{fig:side3}
for the definitions of these variables). Introducing the two
components of the wavevector $(k_x,k_z)$ and  $\delta$, the
attenuation (or evanescence) length, we thus look for solutions as
\begin{eqnarray}
\psi(x,z,t) &\!\! \!\! = \!\!\!\! & \psi_0\, e^{\displaystyle i(k_xx+k_zz-\omega t)}\ e^{\displaystyle-{z'}/{\delta}}\\
&\!\!\!\! =\!\!\!\! &\psi_0\, e^{\displaystyle
(ik_x\!\! +\!\! {\sin\alpha}/{\delta})x\!\! +\!\! (ik_z\!\! -\!\! {\cos\alpha}/{\delta})z\!\! -\!\! i\omega
t},
\end{eqnarray}
since $z'=z\cos\alpha-x\sin\alpha$. Introducing the above ansatz in
Eq.~(\ref{eqhyperbolic}) leads to a complex equation. Separating
real and imaginary parts and defining $\gamma=\omega/N$, we get
\begin{eqnarray}
\delta^2 & = & \frac{\cos^2\alpha}{k_x^2}\ \frac{\gamma^2-\sin^2\alpha}{\gamma^2-1}\\
k_z & = &
k_x\tan\alpha\left(1-{\gamma^{-2}}\right).\label{eq:pourthetaev}
\end{eqnarray}
For $\gamma>1$, we can thus define the angle of propagation of the
evanescent wave $\theta_{ev}=\mbox{atan}({k_x}/{k_z})$.
Equation~(\ref{eq:pourthetaev}) leads to
\begin{equation}
\theta_{ev}(\alpha,\gamma)=\mbox{atan}\left(\frac{\mbox{cotan
}\alpha}{1-{\gamma^{-2}}}\right).\label{thetaev}
\end{equation}
Figure~\ref{fig:theorie} attests that the experimental results
agree very well with formula~(\ref{thetaev}) for run 4 and run 5.
Unexpectedly the case $\theta=19^\circ$ for which the third
harmonic is at the limit of evanescence is however not described
by the above model. Note also that an alternative theoretical description is proposed in Ref.~\cite{TAL05}.

\subsection{Amplitude measurements}
\label{amplitudemeasurements}

 The determination of wave amplitudes is complicated by the
beam-like appearance of the displacement field. Indeed, local
temporal spectral analysis is not sufficient to provide reliable
energy measurements, as phase and group speed are orthogonal.
Spatial integration across the beam are needed to accurately evaluate
the amounts of energy involved in each harmonics.

\begin{figure}[!ht]
\centering
\includegraphics[width=8.5cm]{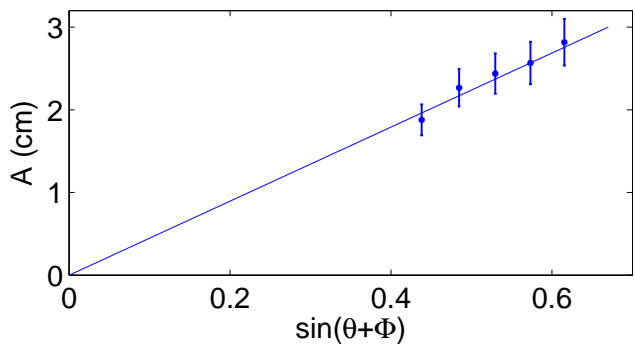}
\caption{Experimental measure of the displacement amplitude of the
emitted wave as a function of the emission angle~$\theta$, for the
five runs of Table~\ref{table:conditions}.}\label{fig:amplitude}
\end{figure}

First to validate the method of measurement, the mean displacement
amplitudes $A=\sqrt{u^2+w^2}/\omega$ of the incident beam is
plotted as a function of the incidence angle. The wavemaker being
tilted at a constant angle $\Phi=13^\circ$ with the horizontal, it
induces a shear motion perpendicular to itself (see
Fig.~\ref{fig:side3}).  The propagating part of this motion is
thus expected to be proportional to $\sin(\theta+\Phi)$.  The very
good agreement shown in Fig.~\ref{fig:amplitude} justifies a
posteriori the method.

\begin{figure}[!ht]
\centering
\includegraphics[width=8.5cm]{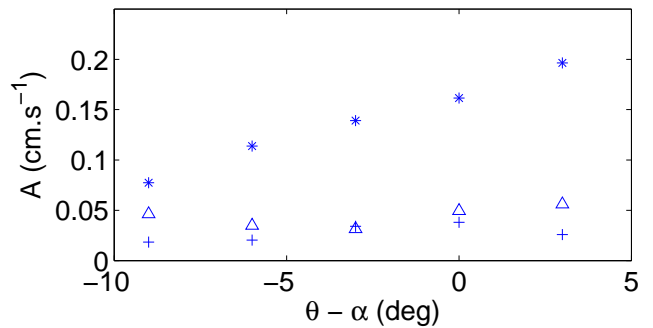}
\caption{Velocity amplitudes of the first three harmonics as
functions of the difference between the emission angle~$\theta$ and the
slope angle $\alpha$, for the five runs of
Table~\ref{table:conditions}. Stars, triangles and crosses
correspond to the first, second and third harmonic
respectively.}\label{fig:amplitudeall}
\end{figure}

Determining the importance of the different harmonics emitted
after the reflection process is an important issue to understand
and hence describe theoretically the reflection process.
After exploring several possibilities, it appears that the best method
to quantitatively characterize the measurements is the following one.
First, it is important to distinguish the two components of velocities vectors,
parallel to the slope and orthogonal to it. The second one vanishes clearly 
close to the slope, satisfying thus the expected boundary conditions.
Away from the slope, because of the possibilities of evanescence it
is very delicate to get reliable amplitudes. On the contrary, the
along slope component of the velocities is easier to deal with. It
has no reason to vanish close to the slope and definitely did not.
We have thus measured the amplitude of the wave by determining the
maximum value of the different harmonics. Results are collected and shown in 
Fig.~\ref{fig:amplitudeall}.  The velocity amplitudes for the
second and third harmonics are 4 times smaller than for the
first harmonic, even for run 4 and run 5 when the third harmonic
is evanescent.  It is important to emphasize that the difference 
between the second and third harmonic is small.

\section{Conclusion}
\label{conclusion}

In this paper, we have reported {\em quantitative} laboratory
measurements of the propagation of internal waves. 
Our experimental set-up produces an incident beam of high quality. Its
large scale allows to reach large Reynolds numbers, such that the
effects of dissipation on the propagation is negligible. This is not
anymore true for the critical reflection mechanism, which involves a
strong reduction of the wavelength, hence increased viscous effects.
Thus, one obtains a steady regime
compatible with the hypothesis of previous theoretical models~\cite{DY99,TAL05}.

The wavevector of the frequency $n$-th harmonics, projected along the slope,
is found to be proportional to
$n$ in the supercritical case. This is in agreement with the theory~\cite{TAL05}.
A different selection mechanism is however observed in the sub-critical
case, for which the wavenumber is equal to the incident one. This is in
contradiction with~\cite{TAL05}, but in agreement with a more simple analysis
previously proposed by Thorpe~\cite{T87}. 

Harmonics with frequency higher than $N$ cannot propagate and remain
trapped near the slope. We document their existence for the first time
and explain their spatial structure.

The long two-dimensional internal waves exciter produces very weak
transversal $y$-variations apparently justifying two-dimensional
predictions in the $(x,z)$-plane. Considering a rotating tank
would require to address the much tougher, fully three dimensional
equation and it is not {\em a priori} clear what will happen. Work
along this line is in progress.

%Fundamental studies of internal waves behavior in the nonlinear
%regime should lead to further improvements in ocean
%modeling~\cite{BMW00}. Indeed, quantitative measurements of
%nonlinear interactions of internal waves with the seafloor may
%help to model small-scale processes, not yet resolved by numerical
%simulations. Besides, the tidal conversion mechanism at a
%submarine ridge is believed to be of paramount
%importance~\cite{PLY05}. Recent three-dimensional numerical
%simulations studies have recently confirmed that internal waves
%are emitted from topography at the single point on the flanks at
%which the topography is equal to the slope of an internal tidal
%beam~\cite{GSBA05,GS05}. Thus, critical reflection of internal
%waves is an interesting fundamental mechanism, with important
%consequences on the parameterization of the boundary layer over
%topography.

\bigskip{\bf Acknowledgments}

We thank G. Delerce and S. Mercier for help during the experiments
and post-processing. Comments to the manuscript by Denis Martinand
are deeply appreciated. This work has been partially supported by
the 2005 PATOM CNRS program and by 2005-ANR project TOPOGI-3D.


\begin{thebibliography}{10}

%\bibitem{BMW00}
%G.~K. Batchelor, H.~K. Moffatt, and M.~G. Worster.
%\newblock {\em {Perspectives in Fluis Dynamics}}.
%\newblock {Cambridge University Press}, 2005.

\bibitem{CW74}
D.~Cacchione and C.~Wunsch.
\newblock Experimental study of internal waves over a slope.
\newblock {\em Journal of Fluid Mechanics}, 66:223, 1974.

\bibitem{DDF04}
T.~Dauxois, A.~Didier, and E.~Falcon.
\newblock Observation of near-critical reflection of internal waves in a stably
  stratified fluid.
\newblock {\em Physics of Fluids}, 16:6, 2004.

\bibitem{DY99}
T.~Dauxois and W.~R. Young.
\newblock Near-critical reflection of internal waves.
\newblock {\em Journal of Fluid Mechanics}, 390:271, 1999.

\bibitem{E85}
C.~C. Eriksen.
\newblock Implications of Ocean Bottom Reflection for Internal Wave Spectra and Mixing.
\newblock {\em Journal of Physical Oceanography}, 15:1145, 1985.

\bibitem{FD00}
A.~Fincham and G.~Delerce.
\newblock Advanced optimization of correlation imaging velocimetry algorithms.
\newblock {\em Experiments in Fluids}, 29:13, 2000.

%\bibitem{GS05}
%T.~Gerkema, C.~Staquet, and P.~Bouruet-Aubertot. 
%\newblock Decay of semi-diurnal internal-tide beams due to subharmonic resonance. 
%\newblock Submitted to {\em  Geophysical Research Letters}, 2005.

%\bibitem{GSBA05}
%T.~Gerkema, C.~Staquet, and P.~Bouruet-Aubertot.
%\newblock Nonlinear effects in internal tide beams and mixing.
%\newblock {\em Ocean Modelling}, 2005.

\bibitem{GGFD05}
L.~Gostiaux, N.~Garnier, E.~Falcon, and T.~Dauxois.
\newblock Generation of internal waves by vibrating cylinders revisited.
\newblock In preparation, 2005.

\bibitem{IN89}
G.~N. Ivey and R.~I. Nokes.
\newblock Vertical mixing due to the breaking of critical internal waves on
  sloping boundaries.
\newblock {\em Journal of Fluid Mechanics}, 204:479, 1989.

\bibitem{IWS00}
G.~N. Ivey, K.~B. Winters, and I.~P. D.~De Silva.
\newblock Turbulent mixing in a sloping benthic boundary layer energized by
  internal waves.
\newblock {\em Journal of Fluid Mechanics}, 418:59, 2000.

\bibitem{JIA99}
A.~Javam, J.~Imberger, and S.~W. Armfield.
\newblock Numerical study of internal wave reflection from sloping boundaries.
\newblock {\em Journal of Fluid Mechanics}, 396:183, 1999.

\bibitem{JIA00}
A.~Javam, J.~Imberger, and S.~W. Armfield.
\newblock Numerical study of internal wave-wave interactions in a stratified
  fluid.
\newblock {\em Journal of Fluid Mechanics}, 415:65, 2000.

\bibitem{M01}
L.~R.~M. Maas.
\newblock Wave focusing and ensuing mean flow due to symmetry breaking in
  rotating fluids.
\newblock {\em Journal of Fluid Mechanics}, 437:13, 2001.

\bibitem{MBSL97}
L.~R.~M. Maas, D.~Benielli, J.~Sommeria, and F.-P.~A. Lam.
\newblock Observation of an internal wave attractor in a confined stably
  stratified fluid.
\newblock {\em Nature}, 388:557, 1997.

\bibitem{ML95}
L.~R.~M. Maas and F.-P.~A. Lam.
\newblock Geometric focusing of internal waves.
\newblock {\em Journal of Fluid Mechanics}, 300:1, 1995.

\bibitem{MK02}
E.~E. McPhee-Shaw and E.~Kunze.
\newblock Boundary layer intrusions from a sloping bottom: A mechanism for
  generating intermediate nepheloid layers.
\newblock {\em Journal of Geophysical Research}, 107:31, 2002.

\bibitem{MR67}
D.~E. Mowbray and B.~S.~H. Rarity.
\newblock A theoretical and experimental investigation of the phase
  configuration of internal waves of small amplitude in a density-stratified
  liquid.
\newblock {\em Journal of Fluid Mechanics}, 28:1, 1967.

\bibitem{PFS05}
A.~Fincham O.~Praud and J.~Sommeria.
\newblock Decaying grid turbulence in a strongly stratified fluid.
\newblock {\em Journal of Fluid Mechanics}, 522:1, 2005.

\bibitem{PT05}
T.~Peacock and A.~Tabei.
\newblock Visualization of nonlinear effects in reflecting internal wave beams.
\newblock {\em Physics of Fluids}, 17:061702, 2005.

%\bibitem{PLY05}
%F.~Petrelis, S.~G.~Llewellyn Smith, and W.~R. Young.
%\newblock Tidal conversion at a submarine ridge.
%\newblock {\em Journal of Physical Oceanography}, 2005.

\bibitem{P77}
O.~M. Philipps.
\newblock {\em The Dynamics of the Upper Ocean}.
\newblock Cambridge University Press, London and New-York, 1977.

\bibitem{SR98}
D.~N. Slinn and J.J. Riley.
\newblock Turbulent dynamics of a critically reflecting internal gravitywave.
\newblock {\em Theoretical and Computational Fluid Dynamics}, 11:281, 1998.

\bibitem{S99}
B.~R. Sutherland.
\newblock Propagation and reflection of internal waves.
\newblock {\em Physics of Fluids}, 11:1081, 1999.

\bibitem{TAL05}
A.~Tabaei, T.~R. Akylas, and K.~Lamb.
\newblock Nonlinear effects in reflecting and colliding internal wave beams.
\newblock {\em Journal of Fluid Mechanics}, 526:217, 2005.

\bibitem{TII97}
S. G. Teoh, G. N. Ivey, and J. Imberger.
\newblock Laboratory study of the interaction between two internal wave rays.
\newblock {\em Journal of Fluid Mechanics}, 336:91, 1997.

\bibitem{T87}
S.~A. Thorpe.
\newblock On the reflection of a train of finite-amplitude internal waves from
  a uniform slope.
\newblock {\em Journal of Fluid Mechanics}, 178:279, 1987.

\bibitem{TH87} S. A. Thorpe and A. P. Haines.
 \newblock
A note on observations of wave reflection on a 20 degree slope. 
\newblock {\em Journal of Fluid Mechanics}, 178:279, 1987.

\bibitem{ZS01} O. Zikanov and D. Slinn.
 \newblock Along-slope current generation by obliquely incident internal waves
\newblock {\em Journal of Fluid Mechanics},  445:235, 2001.

\end{thebibliography}
\end{document}